\documentclass[12pt]{iopart}
\usepackage{amssymb,amsthm}
\usepackage{graphicx}

\newcommand{\beq}[0]{\begin{equation}}
\newcommand{\eeq}[0]{\end{equation}}
\newcommand{\ep}{\varepsilon}
\newcommand{\thet}{\vartheta}
\newcommand{\la}{\langle}
\newcommand{\ra}{\rangle}
\newcommand{\ds}{\displaystyle}
\newcommand{\avh}{\la H_1 \ra}
\newcommand{\pa}{\partial}
\newcommand{\w}{\omega}
\newcommand{\ud}{{\mathrm d}}

\begin{document}
\title[On the stability of multibreathers in Klein-Gordon chains]{On the stability of multibreathers in Klein-Gordon chains}
\author{Vassilis Koukouloyannis$^{1, 2}$ and Panayotis G Kevrekidis$^3$}
\address{$^1$ Department of Physics, Section of Astrophysics, Astronomy and Mechanics, Aristotle University of Thessaloniki, 54124 Thessaloniki, Greece}
\address{$^2$ Department of Civil Engineering, Technological Educational Institute of Serres, 62124 Serres, Greece}
\address{$^3$ Department of Mathematics and Statistics, University of Massachusetts, Amherst MA 01003-9305}
\ead{vkouk@physics.auth.gr}
\date{\today}

\theoremstyle{plain} \newtheorem{theorem}{Theorem}
\theoremstyle{plain} \newtheorem{lemma}{Lemma}
\theoremstyle{definition} \newtheorem{remark}{Remark}

\begin{abstract}
In the present paper, a theorem, which determines the linear stability
of multibreathers excited  over adjacent coupled oscillators in Klein-Gordon chains, is proven. Specifically, it is shown that for soft
nonlinearities, and positive nearest-neighbor 
inter-site coupling, only structures with
adjacent sites excited out-of-phase may be stable, while only in-phase
ones may be stable for negative coupling. The situation is reversed
for hard nonlinearities. This method can be applied to $n$-site 
breathers, where $n$ is any finite number and provides a detailed count 
of the number of real and imaginary characteristic exponents of the breather, 
based on its configuration. In addition, an $\cal{O}(\sqrt{\ep})$ estimation
of these exponents can be extracted through this procedure. To complement 
the analysis, we perform numerical simulations
and establish that the 
results are in excellent agreement with the theoretical predictions, at
least for small values of 
the coupling constant $\ep$. 
\end{abstract}



\maketitle

\section{Introduction}
Intrinsic localized modes (ILMs or discrete breathers) have been a
center of intense theoretical, numerical, as well as experimental
investigations over the past two decades; see e.g. the reviews 
\cite{flawil,macrev,flagor08,aubrev}. Since their theoretical
inception
in the context of anharmonic nonlinear lattices \cite{sietak,pa90} and
subsequent rigorous proof of existence, under appropriate nonresonance
conditions \cite{macaub}, numerous experimental realizations of such
structures have arisen in settings ranging from optical
waveguides
and photorefractive crystals, to micromechanical cantilever arrays
and superconducting Josephson junctions, as well as Bose-Einstein
condensates and electrical lattices, among many others \cite{flagor08}.

While the most fundamental modes among these discrete breathers,
namely
the 1- and 2-site solutions have been analyzed in some detail, much
less is known about the case of 
{\it multi-site breathers} or {\it multibreathers}. The latter were
initially discussed in \cite{macaub}. Since that pioneering work, a
lot of effort 
has been invested in proving the existence of multibreathers in
Klein-Gordon chains (e.g. \cite{ahn1, sepmac, koukicht1}). Regarding the stability of these motions, in \cite{koukichtstab} some
stability results are obtained using the 
formulation of \cite{koukicht1}, but these results are applicable only to few-site excitations. On the other hand, in \cite{arcetal} (see also the more recent discussion of \cite{cuearc})
some theorems about the stability of multibreathers are proven, using
Aubry's band theory \cite{aubrev}, which can be applied to an arbitrary number of site excitations. 

In the present work, a $n$-site breather stability theorem is proven. It generalizes the two previously mentioned works, by proving a detailed 
counting result about the number of real and imaginary characteristic 
exponents of the corresponding breather for arbitrary configurations. 
It should be noted that this result proves a relevant statement made 
in \cite{arcetal} as a claim based in numerical findings. The relevant 
eigenvalues are estimated
to  ${\cal O}(\sqrt{\ep})$.
Our method is based on the notion of the effective
Hamiltonian originally introduced in \cite{ahnmacsep} and generalized
in \cite{mac1, macsep}. This idea has already been used in order to
prove 
existence and stability of multi-site breathers in hexagonal and
honeycomb 
lattices \cite{kouketal, koukmac,koukkour2}. Similar results have been acquired
for the case 
of the discrete nonlinear Schr{\"o}dinger (DNLS) lattice
\cite{pelkevr1,pelkevr2,pelkevr3} and 
were recently used in the study of
discrete solitons 
in hexagonal and honeycomb lattices in \cite{lawetal}.

Our principal result shows that for soft nonlinearities and general,
multi-site
excitations, the relevant structures may only be stable (for positive
values of the coupling) when the adjacent sites are out of phase
by $\pi$ with respect to each other. For negative values of the
coupling, stability is possible for in-phase excitations. This
situation is reversed in the case of hard nonlinearities (i.e.,
in-phase multibreathers are stable for positive weak coupling, while
out-of-phase ones for negative, weak coupling).

Our presentation is organized as follows: in section \ref{system} we
define the system 
under consideration, in section \ref{existence} we set up the general
conditions for 
existence of multibreather solutions, in section \ref{stability} we
acquire the 
theorem about the stability of the previously mentioned solutions and
finally in 
section \ref{numerics} we perform some numerical calculations in order
to verify our 
theoretical predictions.

\section{Definition of the system - Terminology}\label{system}
We define our oscillators by an autonomous Hamiltonian of one
degree of freedom \[H_u=\frac{1}{2}p^2+V(x)\] where $V(x)$ is the
potential function. In this case, the system is integrable since
$H_u$ is always an integral of motion. We assume that $V(x)$
possesses a minimum at $x=0$ (without loss of generality) with $ V''(0)=\w_p^2$ with
$\w_p\in\mathbb{R}$. 

Because of the time reversal symmetry $x(-t)=x(t), p(-t)=-p(t)$ the solution of the oscillator can be written as
\beq x(t)=\sum_{n=0}^{\infty}A_n(J) \cos(nw)\label{fourier}\eeq
where $J, w$ are the action-angle variables. Note that in the action-angle variables the motion of the oscillator is described by
\[\begin{array}{rcl}
w(t)&=&\w t+w_0\\
J(t)&=&\mathrm{const.}
\end{array}\]
where $\w$ is the frequency and $w_0$ is the initial phase of the periodic motion.

We construct our chain by considering a
countable set of oscillators with a
nearest-neighbor coupling through a coupling constant $\ep$. The Hamiltonian then becomes%
\beq H=H_0+\ep
H_1=\sum_{i=-\infty}^{\infty}\left(\frac{1}{2}p_i^2+V(x_i)\right)+\frac{\ep}{2}\sum_{i=-\infty}^{\infty}(x_{i+1}-x_i)^2
\label{hamfull}\eeq
where $x_i$ is the displacement from the equilibrium, and $p_i$ the momentum of the $i$-th oscillator. Note that $H_0$ is trivially
integrable, being separable.

\section{Existence of multibreathers}\label{existence}
Consider the ``anticontinuous'' limit $\ep=0$ where $n+1$ adjacent ``central'' oscillators move in periodic orbits with frequency $\w$ but arbitrary phases, while the the remaining ``non-central'' oscillators lie at rest $(x_i,p_i)=(0,0)$.
This state defines a trivially
localized and time-periodic motion with period $T=2\pi/\w$. We seek conditions
under which this motion can be continued for $\ep\neq0$ to provide a
multibreather of the same frequency $\w$. In the next section we will determine the linear stability of the resulting
solutions. 

We apply the action-angle canonical transformation to the
central oscillators. The system is described now by the set of  variables
$(x_{i},p_{i},w_k,J_k)$ with $k\in \mathbb{S}$ and $i\in
\mathbb{Z}\setminus\mathbb{S}$ where $\mathbb{S}$ is the set of
``central'' oscillators. So the periodic orbit which corresponds to the multibreather
is described at time $t$  by $z(t)=(x_{i}(t),p_{i}(t),w_k(t),I_k(t))$
with $z(t+T)=(x_{i}(t),p_{i}(t),w_k(t)+2\pi,I_k(t))$.

In \cite{ahnmacsep} (extended in \cite{mac1} and \cite{macsep}) it is
proven that under the non-resonance condition $n\w\neq\w_{p}\forall
n\in \mathbb{Z}$ there is an effective Hamiltonian $H^{\mathrm{eff}}$
whose critical points correspond to periodic orbits (in fact, breathers)
of the full system for $\ep$ small enough.
The effective Hamiltonian is defined by
$$H^{\mathrm{eff}}(I_i,A,\phi_i)=\frac{1}{T}\oint H\circ z(t)\
\ud t,$$
where $z$ is a periodic path in the phase space obtained by a continuation
procedure for given relative phases $\phi_i$, relative momenta $I$ and
symplectic ``area'' $A$. In the lowest order of approximation, the
unperturbed orbit $z_0$ can be used instead of $z$.
In our case, this coincides with the averaged Hamiltonian over an angle,
for example $w_0=\w t+w_{0_0}$, due to the linear relationship of $w_0$
with $t$. Since, by construction, the resulting effective Hamiltonian does
not depend on the selected angle $w_0$, and due to the nature of the system, a canonical transformation to the ``central'' oscillators is induced

\beq\begin{array}{lll}
\vartheta=w_0 & &{\cal A}=J_0+\ldots+J_n\\
\phi_1=w_1-w_0& &I_1=J_1+\ldots+J_n\\
\phi_2=w_2-w_1& &I_2=J_2+\ldots+J_n\\
\vdots& &\vdots\\
\phi_{n}=w_n-w_{n-1}& &I_{n}=J_n\\
\end{array}\label{transformation}\eeq
and, in the lowest order of approximation, the effective Hamiltonian
becomes
\beq
H^{\mathrm{eff}}=H_0(I_i)+\ep\avh(\phi_i, I_i)\qquad i=1\ldots n \label{heff}
\eeq
with
\[\avh=\frac{1}{T}\oint H_1 \ud t\]
where the integration is performed along the unperturbed periodic orbit. Note that $\avh$ coincides with $\la H_1\ra_{w_0}$, the average value of $H_1$ over the angle $w_0$ and since $H^{\mathrm{eff}}$ is independent of $\thet$, $A$ is a
constant of motion.

As we have already mentioned, the critical points of this effective
Hamiltonian correspond to breathers. But for non-degenerate critical points, to leading order in $\ep$ this condition reduces to the conditions (which were aquired also in \cite{koukicht1})

\beq\frac{\pa \avh}{\pa \phi_i}=0,\quad \frac{\pa^2\avh}{\pa\phi_i\pa\phi_j}\neq0,\quad \left|\frac{\pa^2H_0}{\pa J_i\pa J_j}\right|\neq0\Rightarrow\frac{\pa \w_i}{\pa J_i}\neq0,\quad \w_p\neq k\w.\label{conditions}\eeq
Note that, since we consider central oscillators with the same frequency, condition (\ref{conditions}c) can be reduced to $\frac{\pa \w}{\pa J}\neq0$. By taking into account (\ref{fourier}) we get (see \ref{avh}
for details)
\beq\avh=-\frac{1}{2}\sum_{m=1}^{\infty}\sum_{i=1}^nA_m^2\cos(m\phi_i)\label{average}\eeq
hence, the condition (\ref{conditions}a) becomes
\beq\sum_{m=1}^\infty mA_m^2\sin(m\phi_i)=0\label{pers_cond}\eeq
which has at least the solutions 
$$\phi_i=0,\pi.$$
Intuitively, we expect that these are the only solutions, in this kind
of systems, however, a general proof of this conjecture is not
presently available \cite{koukkevrroth}.

It is interesting to note in passing here the similarity of the above
conditions
to the Lyapunov-Schmidt persistence conditions obtained in the
context of the DNLS model in
\cite{pelkevr1,pelkevr2,pelkevr3}. However, in the latter case, the
presence of a single frequency enforces the $\phi_i=0,\pi$ condition.

\begin{remark}
Although we don't have a full proof of the above assumption yet, physical considerations suggest its potential validity. For instance, consider
the simplified setting wherein the displacement $x(w)$ (equivalently $x(t)$) from equilibrium is described by the 
truncated series $x(w)=A_0+A_1\cos(w)+A_2\cos(2w)$. Then, the acceleration $a(w) \equiv \ddot{x}(w)$ reads $a(w)=-\w^2\left[A_1\cos(w)+4A_2\cos(2w)\right]$. Since we know that $a$ in the two edges of the motion should be $a(0)<0$ and $a(\pi)>0$, this means that $A_1+4A_2>0$ and $A_1-4A_2>0$ and finally $A_1^2>16A_2^2$. Since for this case (\ref{pers_cond}) reads $A_1^2 \sin(\phi) + A_2^2 \sin(2 \phi)=\sin(\phi) [A_1^2 +4 A_2^2 \cos(\phi)]=0$,
we conclude that in this special case, the above physical considerations preclude solutions other than
$\phi=0, \pi$.

\end{remark}

\section{Stability of the multibreather solutions}\label{stability}
The linear stability of the fixed point of $H^{\mathrm{eff}}$ determines
also the linear stability of the breather. This is proven in
\cite{ahnmacsep} for the first order approximation to $H^{\mathrm{eff}}$,
under the assumption of distinct eigenvalues of the first order matrix,
and in \cite{mac1} for the general case. This fact has already been
used in order to study the stability of 3-site breathers in
\cite{koukmac}. Again, there is a direct analog of this in the DNLS
case,
whereby the Jacobian of the Lyapunov-Schmidt conditions in
\cite{pelkevr1,pelkevr2,pelkevr3} is, to leading order, directly analogous to the
squared eigenvalues of the full linearization problem.

To make things more precise, the linear stability of a multibreather,
is determined by its {\it Floquet multipliers} (see
e.g. \cite{aubrev}), which are the eigenvalues of the monodromy matrix
of the corresponding periodic orbit. If {\it all} the multipliers lie
on the unit circle the breather is linearly stable, otherwise the
breather is unstable. Due to the Hamiltonian character of the system
if $\lambda$ is a multiplier so are $\lambda^*, \lambda^{-1},
{\lambda^*}^{-1}$. In particular, for multibreathers, when $\ep=0$ all
the multipliers lie in two conjugate bundles at $e^{\pm i\w_pT_b}$
except for $n+1$ pairs, which lie at unity and correspond to the
central oscillators. When this solution is continued for $\ep\neq0$ the
multipliers which belong to the two bundles, being of the same Krein
kind (e.g. \cite{yac}),  move along the unit circle to form the phonon
band. On the other hand, one pair of the multipliers of the central
oscillators will remain at $1$ because of the corresponding invariance of the system
while the rest can move either along the unit circle or 
outside the unit circle determining in 
this way the linear stability of the multibreather.

We define the {\it characteristic exponents} $\sigma_i$ of the multibreather, or equivalently of the corresponding periodic orbit, as
$$\lambda_i=e^{\sigma_i T_b}.$$
The non-zero characteristic exponents
of the central oscillators correspond to the 
eigenvalues of the $(2n\times 2n)$ {\it stability matrix}
\cite{ahnmacsep, mac1} 
${\bf E}={\bf \Omega}
D^2H^{\mathrm{eff}}$, where
${\bf \Omega}=\left(\begin{array}{cc}\bf O&-\bf I\\\bf I&\bf O\end{array}\right)$ and $\bf I$ the
$n\times n$ identity matrix. According to the above, for linear stability we demand that all the eigenvalues of $\bf E$ be purely imaginary.
The stability matrix $\bf E$, to leading order of approximation and by
taking into
 consideration (\ref{heff}), becomes
\beq\fl \bf E=\left(\begin{array}{c|c}\bf A&\bf B\\ \hline\bf C&\bf D\end{array}\right)=\left(\begin{array}{c|c}\ep\bf A_1&\ep\bf B_1\\ \hline\bf C_0+\ep\bf C_1&\ep\bf D_1\end{array}\right)=\left(\begin{array}{c|c}
-\ep\ds\frac{\pa^2\avh}{\pa\phi_i\pa I_j}&-\ep\ds\frac{\pa^2\avh}{\pa\phi_i\pa\phi_j}\\[10pt]
\hline\\[-8pt]
\ds\frac{\pa^2H_0}{\pa I_i I_j}+\ds\ep\frac{\pa^2\avh}{\pa I_i\pa I_j}&\ds\ep\frac{\pa^2\avh}{\pa\phi_j\pa I_i}
\end{array}\right).\label{e1}\eeq
Using (\ref{average}), the elements of $\bf E$ take the form

$$\frac{\pa^2\avh}{\pa\phi_i\pa I_j}=\sum_{m=1}^{\infty}mg(J)\sin(m\phi_i)$$
with $\quad\ds g(J)=\frac{\pa}{\pa I_j}\left(A_m(J_{i-1})A_m(J_{i})\right)\big|_{J_{i-1}=J_{i}=J}\quad$ and $\quad i,j=1\ldots n,$\\
while, (\ref{apb}), 
$$\begin{array}{cc}\ds\frac{\pa^2H_0}{\pa I_i\pa I_{j}}=&\left\{\begin{array}{cl}
\ds2\frac{\pa\w}{\pa J}&j=i\\[8pt]
\ds-\frac{\pa\w}{\pa J}&j=i\pm1\\[8pt]
\ds 0& \mathrm{else}
\end{array}\right.\end{array}\ \quad and 
\qquad
\begin{array}{cc}\ds\frac{\pa^2\avh}{\pa \phi_i\pa \phi_j}=&\left\{\begin{array}{cl}
\ds f(\phi_i)&j=i\\[8pt]
0&j\neq i
\end{array}\right.\end{array}
$$

where
\beq f(\phi)=\frac{1}{2}\sum_{n=1}^{\infty}n^2A_n^2\cos(n\phi)\label{f}.\eeq
Since we consider solutions with\ $\phi_i=0, \pi$, where $\bf A=\bf
D=\bf O$, 
 the stability matrix of (\ref{e1}) becomes 
\beq\bf E=\left(\begin{array}{cc}\bf O&\bf B\\\bf C&\bf O\end{array}\right)=\left(\begin{array}{cc}\bf O&\ep\bf B_1\\\bf C_0+\ep \bf C_1&\bf O\end{array}\right)\label{e2}\eeq
Due to Lemma \ref{lem_oe2}, the $\frac{\pa^2\avh}{\pa I_i\pa I_j}$
terms contribute only to higher (than the leading) order, and hence
will not be considered further in what follows.
\begin{lemma}\label{lem_oe2}
The leading order of approximation of the eigenvalues of $\bf E$ is
${\cal O}(\sqrt{\ep})$. The term ${\bf C_1}=\frac{\pa^2\avh}{\pa I_i\pa
  I_j}$ in (\ref{e2}) only affects the eigenvalues at ${\cal O}(\ep^{3/2})$. 
\end{lemma}
The proof can be found in \ref{oe2}.\qed
\\ $ $\\
As it is also shown in \ref{oe2}, up to the leading order of approximation, we have
\beq\sigma_{\pm i}=\pm\sqrt{\ep\,\chi_{1i}}+{\cal O}(\ep^{3/2})\quad i=1\ldots n\label{sx},\eeq
where $\sigma_{\pm i}$ (the characteristic exponents) are the eigenvalues of $\bf E$ and $\chi_{1i}$ are the eigenvalues of $\bf B_1C_0$. So, the sign of $\chi_{1i}$ define the stability of the multibreather. Let, $f_i=f(\phi_i)$. Then, $\bf B_1C_0$ becomes
\beq{\bf B_1}\cdot{\bf C_0}=-\frac{\pa\w}{\pa J}{\bf Z}=-\frac{\pa\w}{\pa J}\left(\begin{array}{ccccc}
2f_1&-f_1&0 & & \\
-f_2&2f_2&-f_2&0 & \\
 &\ddots&\ddots&\ddots & \\
 & 0 &-f_{n-1}&2f_{n-1}&-f_{n-1}\\
 &  &0 & -f_n&2f_n
\end{array}\right)\label{z}\eeq

\begin{lemma}\label{lem2}
Let $z_i$ be the eigenvalues of $\bf Z$. Then, the number of positive $z_i$'s equals the number of positive $f_i$'s, while the number of negative $z_i$'s equals the number of negative $f_i$'s.
\end{lemma}
The proof can be found in \ref{ap_sign}.\qed

\begin{lemma}\label{lem3}
Assuming the absence of solutions of (\ref{conditions}a) other than $\phi_i=0,
\pi$, 
then $f(0)>0$ and $f(\pi)<0$.
\end{lemma}
\begin{proof}
The fact that $f(0)>0$ is obvious from (\ref{f}) since $A_i$ are the Fourier coefficients of a smooth real function. On the other hand, 
\[F(\phi)=-\frac{1}{2}\sum_{n=1}^{\infty}A_n^2\cos(n\phi)\]
is a continuous function. Since the values $\phi=0, \pi$ correspond to
the extrema of
 $F(\phi)$, because of continuity, one of them corresponds to a local
 minimum while the other 
corresponds to a local maximum. So, since
$f(0)=\frac{\ud^2F(\phi)}{\ud \phi^2}\Big|_{\phi=0}>0$ corresponds to
the minimum, therefore $f(\phi)=\pi$ must correspond to the maximum of
$F(\phi)$ and 
$f(\pi)=\frac{\ud^2F(\phi)}{\ud \phi^2}\Big|_{\phi=\pi}<0$.
\end{proof}

\begin{lemma}\label{lem4}
If $\ep\frac{\pa \w}{\pa J}<0$ and $\phi_i=\pi\quad\forall i=1\ldots n$, or if $\ep\frac{\pa \w}{\pa J}>0$ and $\phi_i=0\quad\forall i=1\ldots n$, then all the eigenvalues of $E$ are purely imaginary up to ${\cal O}(\sqrt{\ep})$ terms.  
\end{lemma}
\begin{proof}
Due to (\ref{z}), we have $\chi_{1i}=-\frac{\pa \w}{\pa J}z_i$. So, by using (\ref{sx}) we get
\beq\sigma_{\pm i}=\pm\sqrt{-\ep\, \frac{\pa \w}{\pa J}\, z_{i}}+{\cal O}(\ep^{3/2}).\label{s_approx}\eeq
The sign of $z_i$ is defined by the value of $\phi_i$ according to
lemmas \ref{lem2} and \ref{lem3}, which, in turn, completes the proof
of the lemma.
\end{proof}

If the eigenvalues $\lambda_i$ are imaginary and distinct up to ${\cal
  O}(\sqrt{\ep})$ terms the higher order terms cannot push them outside
the imaginary axis for a variance of $\ep$, say $\Delta\ep$, small
enough, because of continuity. On the other hand, if the eigenvalues
$\lambda_i$ have multiplicity $>1$ up to ${\cal O}(\sqrt{\ep})$ terms
the higher order terms can, in principle, push them outside the
imaginary axis for $\Delta\ep$ arbitrary small, which would cause
complex instability, through a Hamiltonian Hopf bifurcation. 
This, however, cannot happen in our system since a specific symplectic signature property holds.

\begin{lemma}\label{lem5}
If the eigenvalues of $\bf E$ are imaginary up to ${\cal O}(\sqrt{\ep})$ terms they remain imaginary up to all orders of approximation.
\end{lemma}
\begin{proof}
If the eigenvalues of $\bf E$ are imaginary up to some order of approximation, then, according to \cite{mac2}, if the corresponding quadratic form of $D^2H^{\mathrm{eff}}$ is definite, then the eigenvalues remain imaginary for all orders 
of approximation. The matrix $D^2H^{\mathrm{eff}}$ is
$$D^2H^{\mathrm{eff}}=\left(\begin{array}{cc}
\ds\frac{\pa^2 H_0}{\pa J_i\pa J_j}& O\\
O&\ds\frac{\pa^2\avh}{\pa\phi_i\pa\phi_j}
\end{array}\right)$$
and the corresponding quadratic form is $\delta^2H^{\mathrm{eff}}=(\overline{I}, \overline{\phi})\cdot D^2H^\mathrm{eff}\cdot(\overline{I}, \overline{\phi})^T$, with $\overline{I}=(I_1,\ldots, I_n)$ and $\overline{\phi}=(\phi_1,\ldots, \phi_n)$. Finally we get
$$\delta^2H^{\mathrm{eff}}=\frac{\pa\w}{\pa J}\left[I_1^2+(I_2-I_1)^2+\ldots+(I_{n}-I_{n-1})^2+I_n^2\right]+\ep\left[ f(\phi_1)\phi_1^2+\ldots+f(\phi_n)\phi_n^2\right].$$
This quadratic form remains definite for all the configurations which
are described in Lemma \ref{lem4}. So, even in the case of higher
multiplicity, 
the imaginary eigenvalues of $E$ remain on the imaginary axis.
\end{proof}
The sequence of the above lemmas leads to our main stability theorem,
as follows:
\begin{theorem}\label{thm1}
Under the assumption that (\ref{pers_cond}) has no other solutions than $\phi_i=0, \pi$, then, if $\ep\frac{\pa \w}{\pa J}<0$ the only configuration which leads to
linearly stable multibreathers, for $|\ep|$ small enough, is the one
with $\phi_i=\pi\quad\forall i=1\ldots n$ (out-of-phase
multibreather), while if $\ep\frac{\pa \w}{\pa J}>0$ the only linearly
stable configuration, for $|\ep|$ small enough, is the one with
$\phi_i=0\quad\forall i=1\ldots n$ (in-phase multibreather).
Moreover, for $\ep\frac{\pa \w}{\pa J}<0$ (respectively, $\ep\frac{\pa
  \w}{\pa J}>0$), for unstable configurations, their number of 
unstable eigenvalues will be precisely equal to the number of nearest
neighbors
which are in (respectively, out of) phase between them.
\end{theorem}
\begin{proof}
Since for a linear stable multibreather we need imaginary eigenvalues of $E$, the only possible configurations for stability are the ones described by the theorem, as it can be shown from lemmas \ref{lem3} and \ref{lem4}. The multibreather will remain stable for small enough values of $|\ep|$, until the eigenvalues which correspond to the central oscillators will collide with the linear spectrum, causing a Hamiltonian Hopf bifurcation, leading to complex instability.  
\end{proof}
We note in passing that the above theorem bears a direct analogy to
Theorem 3.6 of \cite{pelkevr1} for the DNLS case. 
\begin{remark}
Note that if the the on-site potential is even $V(x)=V(-x)$ then the cosine Fourier series of $x(t)$ becomes
$$x(t)=A_0+\sum_{n=1}^\infty A_{2n-1}\cos[(2n-1)w]$$
and $f(\phi)$ becomes
$$f(\phi)=\frac{1}{2}\sum_{n=1}^\infty (2n-1)^2A_{2n-1}^2\cos[(2n-1)\phi]$$
which means $f(\pi)=-\frac{1}{2}\sum_{n=1}^\infty(2n-1)^2A_{2n-1}^2<0$. So, Theorem \ref{thm1} can be reformulated without the need of exclusion of possible other solutions of  (\ref{pers_cond}).
\end{remark}

\section{Numerical Results}\label{numerics}

\begin{figure}[!ht]
	\centering
	\begin{tabular}{ccc}
		\includegraphics[height=4cm]{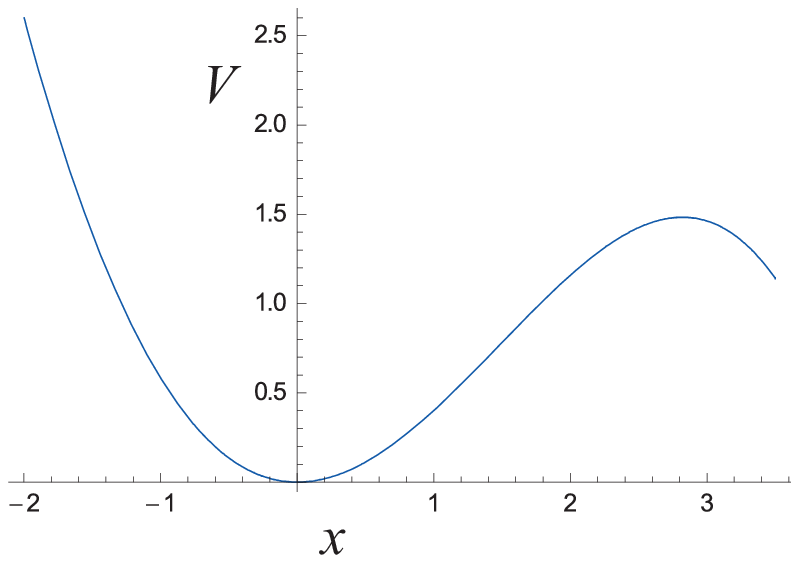}&\hspace{0.5cm} &\includegraphics[height=4cm]{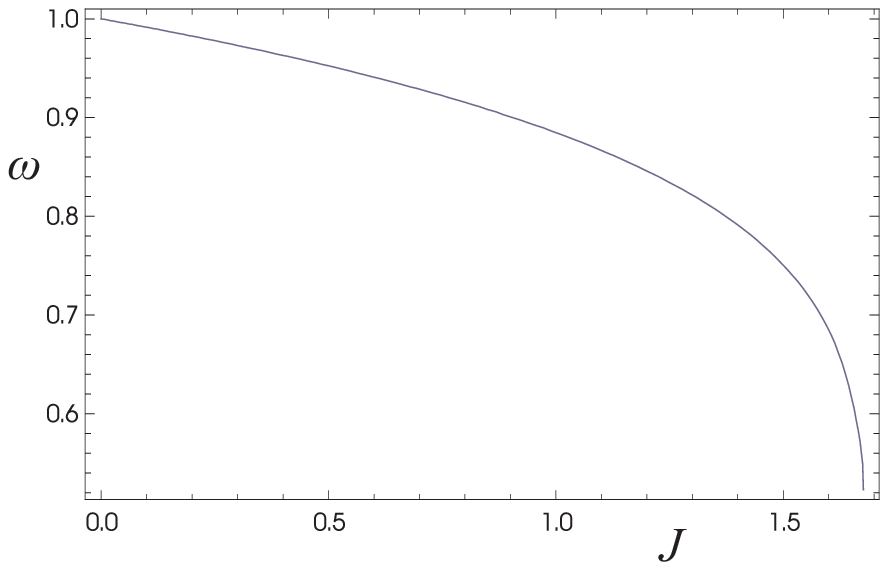}\\
		(a) &\hspace{0.5cm} & (b)
	\end{tabular}	
	\caption{Panel (b) shows the dependence of $\w$ with respect
          to  $J$ 
for the specific $V(x)$ of panel (a).}
	\label{fig:dwdJ}
\end{figure}

As a prototypical numerical demonstration, consider a chain consisting
of oscillators with on-site quartic potential
$V(x)=\frac{x^2}{2}-0.27\frac{x^3}{3}-0.03\frac{x^4}{4}$. This
potential is softening $\left(\frac{\pa\w}{\pa J}<0\right)$ as it can
be seen in figure \ref{fig:dwdJ}. We will consider the orbit with period $T=\frac{2\pi}{\w}=7.434$ which corresponds to amplitude of oscillation $x_{max}=1.949275\Rightarrow J=1.20306\Rightarrow\frac{\pa\w}{\pa J}=-0.224556$. For the same orbit we get $f(0)=1.423404$ and $f(\pi)=-1.279544$.
 
\subsection{2-site multibreathers}
\begin{figure}[!htbp]
\centering
	\begin{tabular}{ccc}
		\includegraphics[height=4cm]{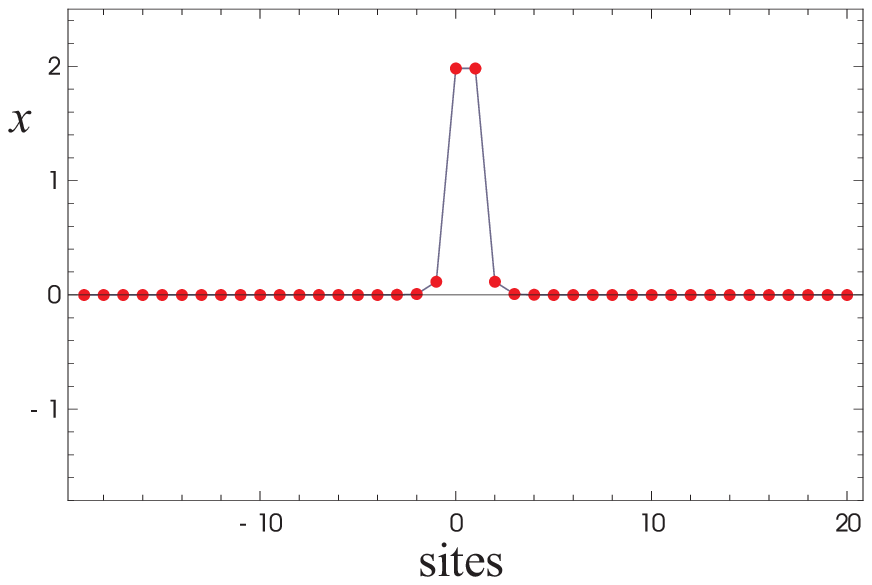}& &\includegraphics[height=4cm]{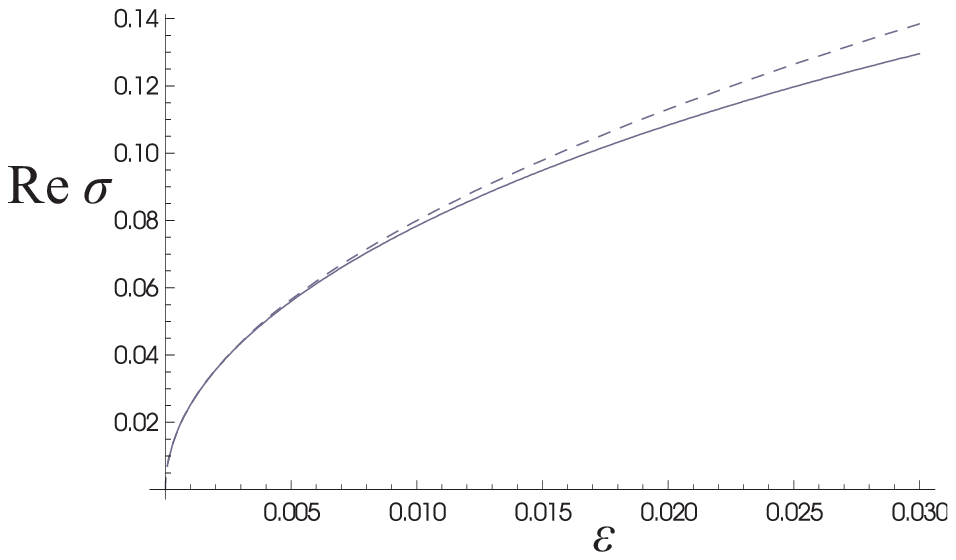}\\
		(a)& &(b)
	\end{tabular}
	\caption{(a) The profile of an in-phase 2-site breather for $\ep=0.02$. In
          (b) the real part of 
$\sigma_1$, for increasing values of $\ep$, is shown. The solid line
represents the 
numerically calculated value, while the dashed one is the one
resulting 
from (\ref{s1_2site}). }
	\label{fig:2site_in}
\end{figure}

\begin{figure}[!htbp]
\centering
	\begin{tabular}{ccc}
		\includegraphics[height=4cm]{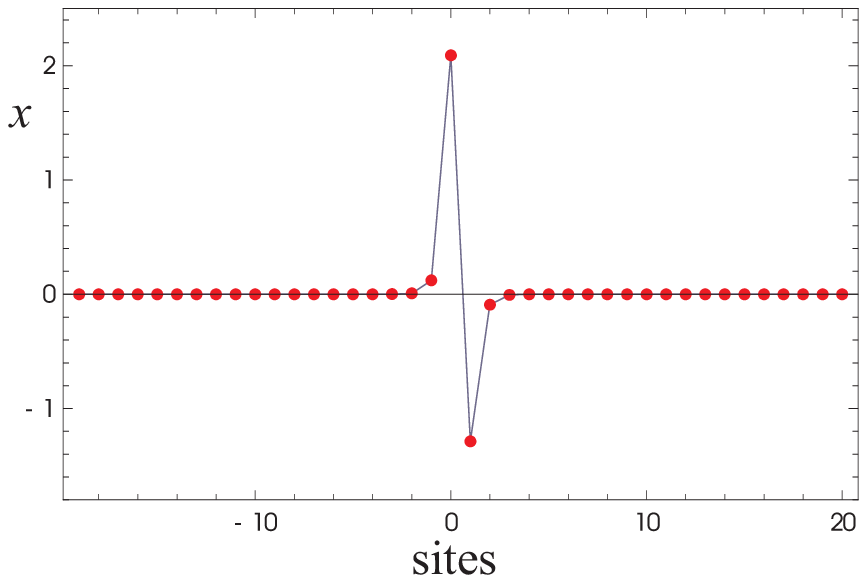}&\hspace{0.5cm} &\includegraphics[height=4cm]{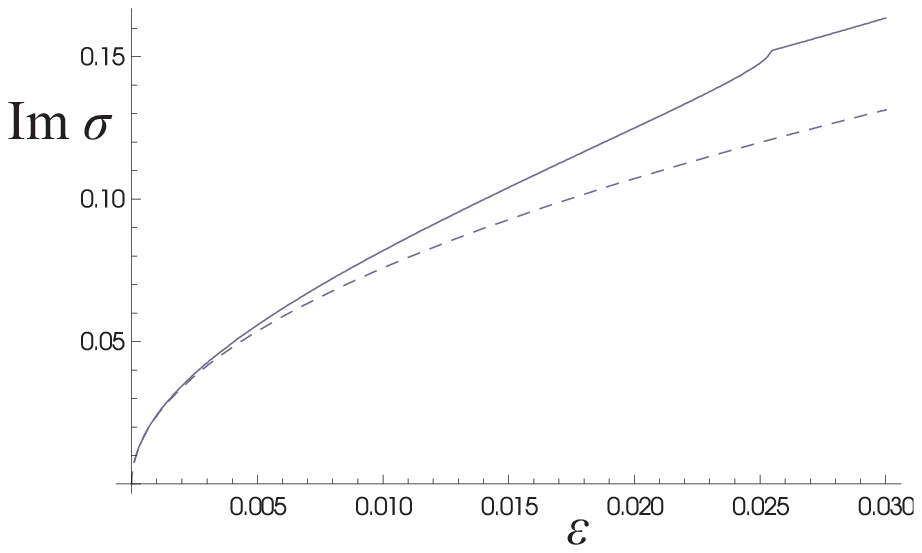}\\
		(a)&\hspace{0.5cm} &(b)
		\end{tabular}
\caption{(a) The profile of an out of phase 2-site
          breather for $\ep=0.02$. In (b) the imaginary
 part of $\sigma_1$, for increasing values of $\ep$, is shown. The
 solid line 
represents the numerically calculated value, while the dashed one is
the one 
resulting from (\ref{s1_2site}).}
	\label{fig:2site_out}
\end{figure}

We consider first the case of two cental oscillators (two oscillators
moving at the anti-continuous limit). In this case there is only one
$\phi=w_2-w_1=w_{20}-w_{10}$ and consequently only 
a pair of characteristic exponents. The leading order approximation of $\sigma_i$ is, according to (\ref{s_approx}), 
\beq\sigma_{\pm 1}=\pm\sqrt{2\ep\frac{\pa\w}{\pa J}f(\phi)}\label{s1_2site}.\eeq
The resulting 2-breathers are: 
\begin{itemize}
 \item The {\it in-phase} 2-breather with $\phi=0$. In our example
   where $\ep>0$ and $\frac{\pa \w}{\pa J}<0$ we get $\sigma_{\pm 1}\in
   \mathbb{R}$, which leads to an {\it unstable} 2-breather. In
   figure \ref{fig:2site_in}a, the 
profile of the in-phase 2-breather is shown while in
figure \ref{fig:2site_in}b the real 
part of the positive characteristic exponent of the central oscillator
$\sigma_1$ as 
calculated by the numerical simulation is shown (solid line) together
with the 
theoretical $\cal{O}(\sqrt{\ep})$ prediction of $\sigma_1$ in
(\ref{s1_2site}) (dashed line). 
We can see that for small values of $\ep$ the agreement is excellent,
while for larger 
values of $\ep$, where the higher order terms of $\sigma_1$ become
significant, the two lines
start to diverge.
 \item The {\it out of phase} 2-breather with $\phi=\pi$
   (see figure \ref{fig:2site_out}). In our 
example it is $\sigma_{\pm 1}\in \mathbb{I}$, which leads to a {\it
  linearly stable} 2-breather. 
In figure \ref{fig:2site_out}a, the profile of the out of phase
2-breather is shown, while in 
figure \ref{fig:2site_in}b the imaginary part of $\sigma_1$ is
shown. The solid line represents 
the numerically calculated value while the dashed line is the
theoretically approximated value. 
For small values of $\ep$ the two lines again nearly coincide, while
for larger values of 
$\ep$, the two lines diverge. For $\ep\simeq0.0254$ the solid line
possesses a 
cusp which results from the collision of $\sigma_1$ with the linear
spectrum. At this point two characteristic exponents (and their
  conjugates) acquire a nonzero real part, through a Hamiltonian Hopf
  bifurcation, and the corresponding multibreather becomes
  unstable. This is the typical mechanism through which 
multibreather solutions
identified herein as stable for small $\ep$ eventually become
unstable as the coupling is increased.
 \end{itemize}

\subsection{3-site multibreathers}
The next step is to consider three central oscillators. In this case there exist two independent $\phi_i$'s, $\phi_1$ and $\phi_2$. So, there are three 
relevant configurations to examine, which correspond to the three possible 
combinations of $\phi_i$.
\begin{itemize}
\item $\phi_1=\phi_2=0$ {\it (In-phase multibreather)}. Following (\ref{s_approx}), the leading
  order 
approximation of the four characteristic exponents of the 3-breather is 
\beq\sigma_{\pm 1}=\pm \sqrt{-\ep\frac{\pa \w}{\pa J}f(0)}\ \ ,\quad
\sigma_{\pm 2}=
\pm \sqrt{-3\ep\frac{\pa \w}{\pa J}f(0)}.\label{s_in_3site}\eeq
In our example this is an unstable configuration since $\sigma_{\pm1,
  \pm2}\in \mathbb{R}$. 
In figure \ref{fig:3site_in}a the profile of a 3-site in-phase breather for $\ep=0.02$
is shown, while in 
figure \ref{fig:3site_in}b the real part of the corresponding
characteristic 
exponents $\sigma_{1,2}$ is shown. Again, the solid line denotes the
numerically calculated 
values and the dashed ones represent the theoretical predictions. The
agreement is very 
good, especially for small values of $\ep$, illustrating the accuracy
of
our theoretical predictions.
\begin{figure}[!htbp]
\centering
	\begin{tabular}{ccc}
		\includegraphics[height=4cm]{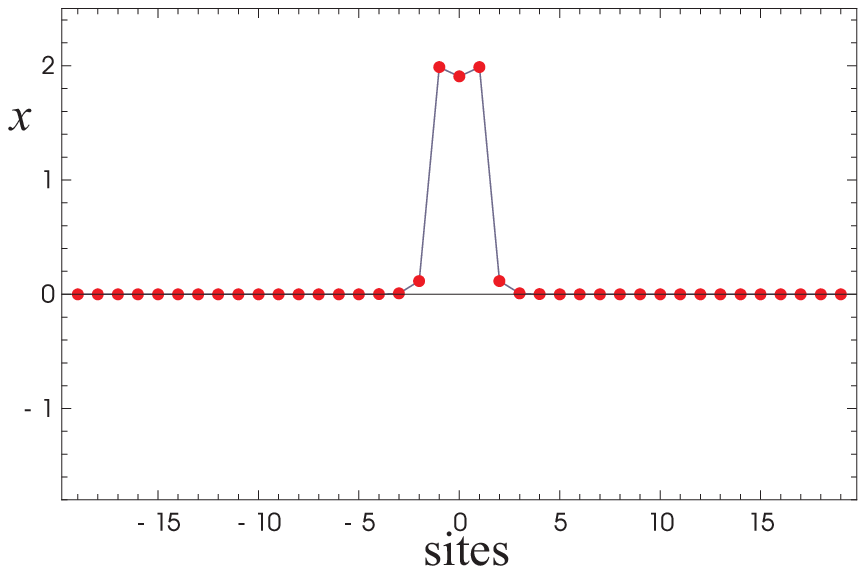}&\hspace{0.5cm} &\includegraphics[height=4cm]{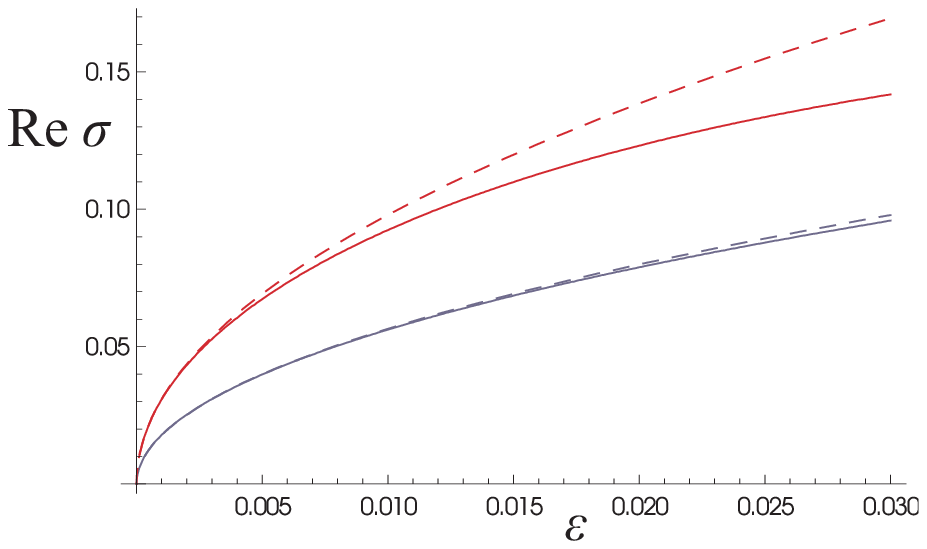}\\
		(a)&\hspace{0.5cm} &(b)
		\end{tabular}
	\caption{(a) The profile of an in-phase 3-site breather for $\ep=0.02$. In
          (b) the real part of $\sigma_{1,2}$, for increasing values
          of $\ep$, is shown. The solid lines represent the numerically
          calculated values, while the dashed lines correspond to the
          theoretical
result of (\ref{s_in_3site}).}
	\label{fig:3site_in}
\end{figure}

\item $\phi_1=\phi_2=\pi$ {\it (Out-of-phase multibreather)}. 
In this case, the leading order approximation of the corresponding characteristic exponents is
\beq\sigma_{\pm1}=\pm \sqrt{-\ep\frac{\pa \w}{\pa J}f(\pi)}\ \ ,\quad\sigma_{\pm2}=\pm \sqrt{-3\ep\frac{\pa \w}{\pa J}f(\pi)}.\label{s_out_3site}\eeq
In our example this is a stable configuration since $\sigma_{\pm1},
\sigma_{\pm2}\in \mathbb{I}$. In figure \ref{fig:3site_out}a the profile
of a 3-site out of phase breather is shown, while in
figure \ref{fig:3site_in}b, the imaginary part of the corresponding
characteristic exponents is shown. Once again, the agreement 
between the two
lines can be noted, at least for small $\ep$. For $\ep\simeq0.019$, the
line which corresponds 
to $\sigma_2$ appears to change slope, a feature which is due to its
collision
with the multipliers stemming  from the phonon band.  For larger values of $\ep$ the multibreather is unstable.

\begin{figure}[!htbp]
\centering
	\begin{tabular}{ccc}
		\includegraphics[height=4cm]{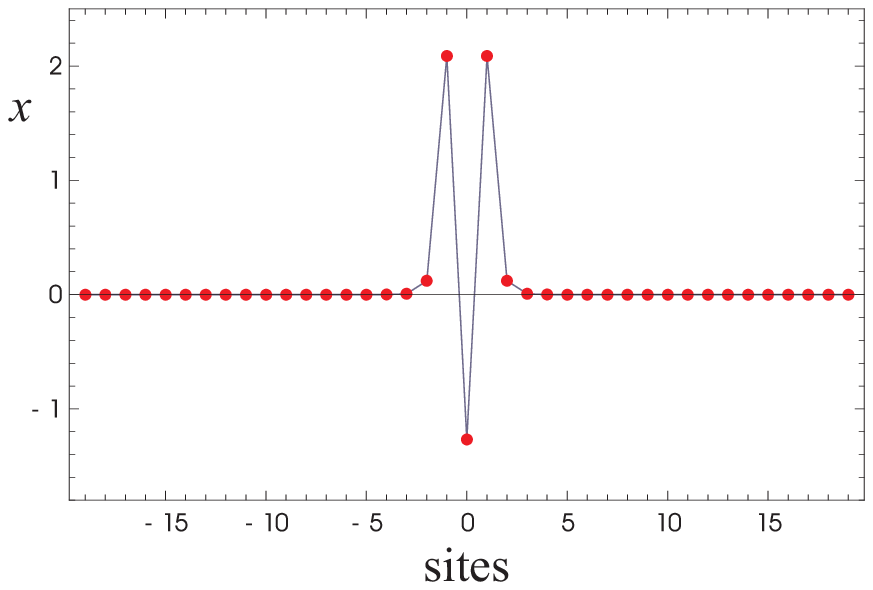}&\hspace{0.5cm}&\includegraphics[height=4cm]{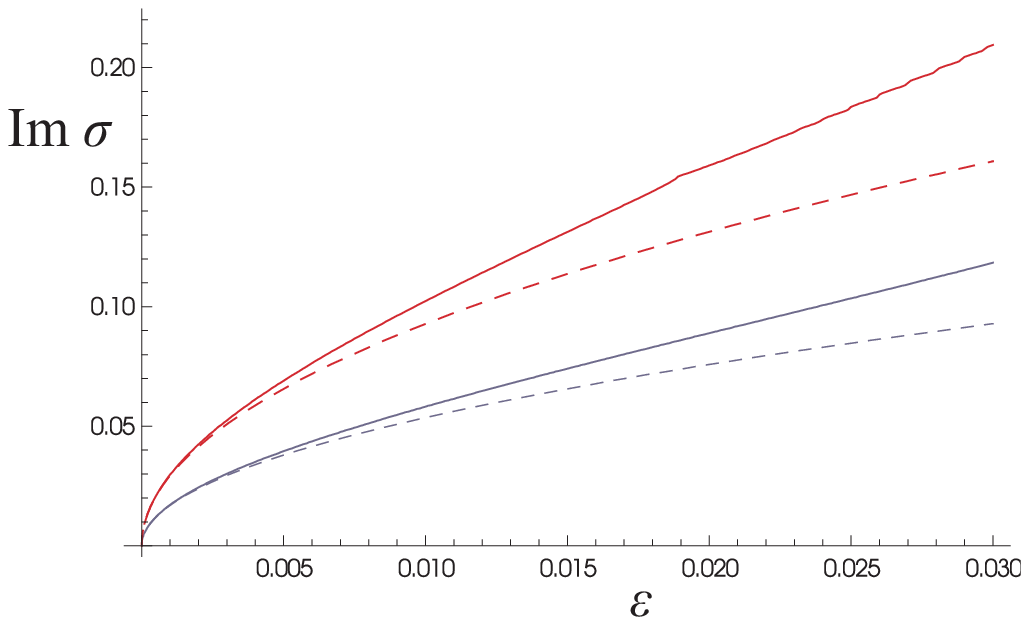}\\
		(a)&\hspace{0.5cm} &(b)
	\end{tabular}	
	\caption{(a) The profile of an out-of-phase 3-site breather for $\ep=0.02$. In (b) the imaginary part of $\sigma_{1,2}$, for increasing values of $\ep$, is shown. The solid lines represent the numerically calculated values, while the dashed ones are the ones resulting from (\ref{s_out_3site}).}
	\label{fig:3site_out}
\end{figure}

\item $\phi_1=0, \phi_2=\pi$. 
In this case, the corresponding leading order approximation of the characteristic exponents is
\beq\begin{array}{ll}\sigma_{\pm1}=\pm \sqrt{-\ep\frac{\pa \w}{\pa J}\left(f_1+f_2-\sqrt{f_1^2+f_2^2-f_1f_2}\right)}\\ 
\sigma_{\pm2}=\pm \sqrt{-\ep\frac{\pa \w}{\pa J}\left(f_1+f_2+\sqrt{f_1^2+f_2^2-f_1f_2}\right)}\end{array}\label{s_in_out_3site}\eeq
with $f_1=f(0)$ and $f_2=f(\pi)$. In our example,
$\sigma_{\pm1}\in\mathbb{I}$ and $\sigma_{\pm2}\in\mathbb{R}$
[it is straightforward to show that this will always be the
case if $f_1 f_2 < 0$], so the
corresponding configuration, shown in figure \ref{fig:3site_in_out}, is
unstable. 
In figures \ref{fig:3site_in_out}(b) and \ref{fig:3site_in_out}(c), the
imaginary and real
 parts of $\sigma_1$ and $\sigma_2$ are shown. Note that for
 $\ep\simeq0.03$, $\sigma_1$ 
enters the phonon band as can be seen in
figure \ref{fig:3site_in_out}(c), where its real 
part becomes nonzero and the corresponding multibreather becomes unstable.
\begin{figure}[!htbp]
	\begin{tabular}{ccc}		\hspace{-1.5cm}\includegraphics[height=3.5cm]{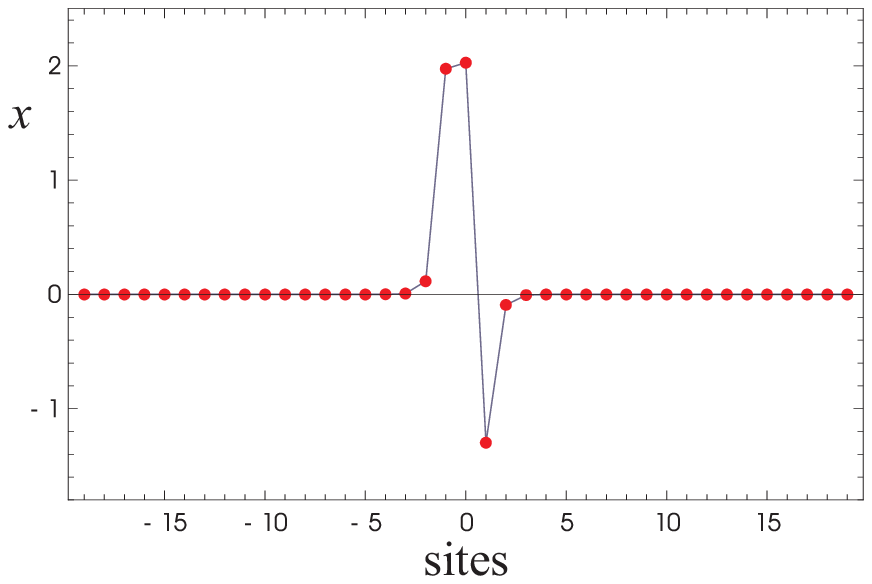}&\includegraphics[height=3.5cm]{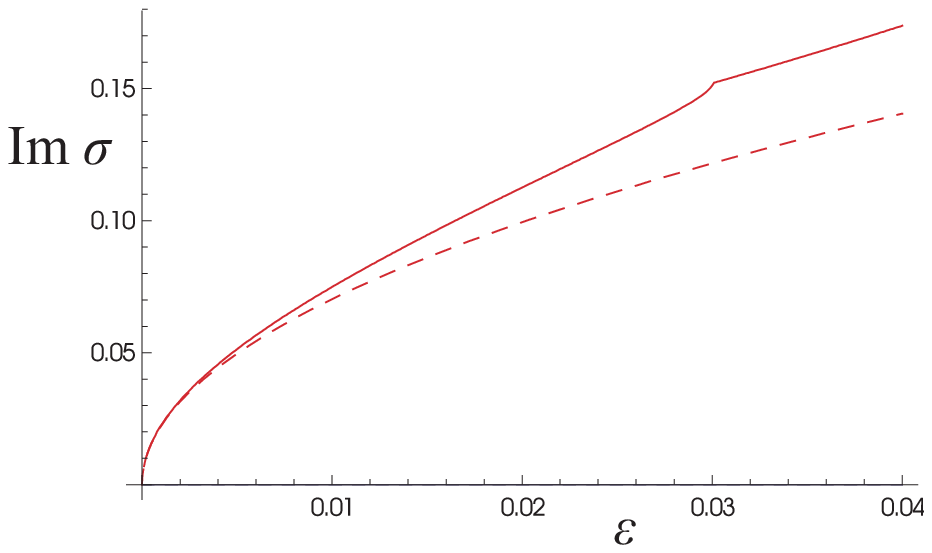}&\includegraphics[height=3.5cm]{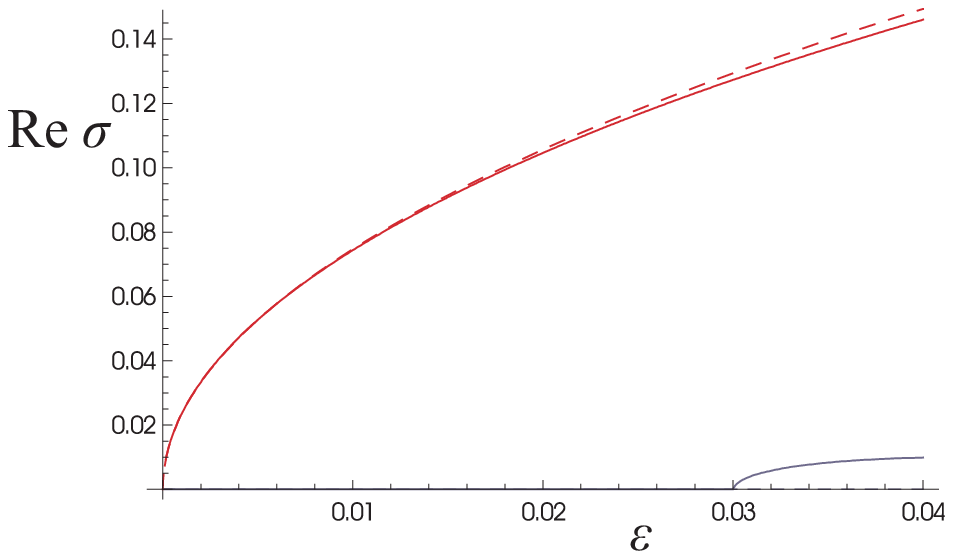}\\\hspace{-1.5cm}(a)&(b)&(c)
\end{tabular}
	\caption{Panel (a) shows the profile of the multibreather
          corresponding to the 
$\phi_1=0, \phi_2=\pi$ configuration for $\ep=0.02$. In (b) and (c) the imaginary and
the real parts 
of $\sigma_{1,2}$, for increasing values of $\ep$, are shown
correspondingly. 
The solid lines represent the numerically calculated values, while the
dashed ones are the ones resulting from (\ref{s_in_out_3site}).}
	\label{fig:3site_in_out}
\end{figure}
\end{itemize}

\subsection{5-site multibreathers}
\begin{figure}[!htbp]
	\begin{tabular}{ccc}	\hspace{-1.5cm}\includegraphics[height=3.5cm]{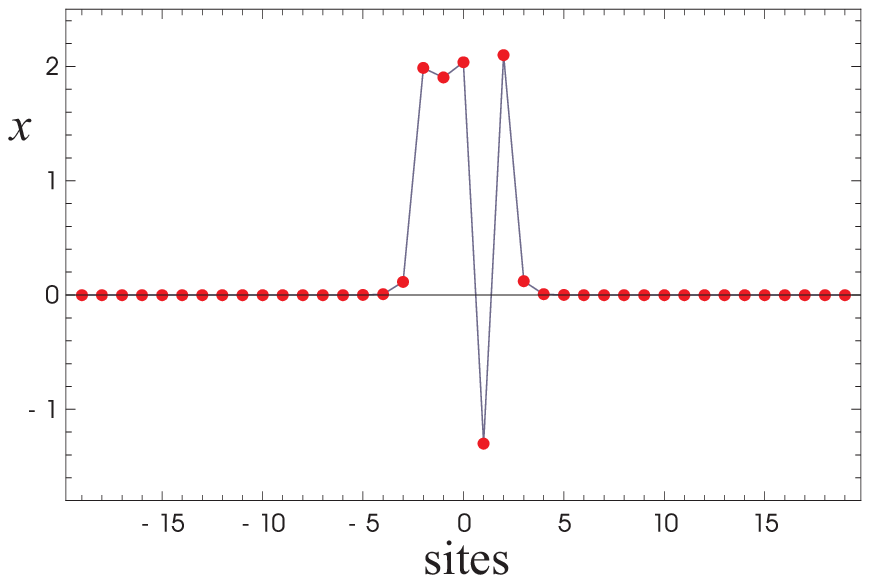}&\includegraphics[height=3.5cm]{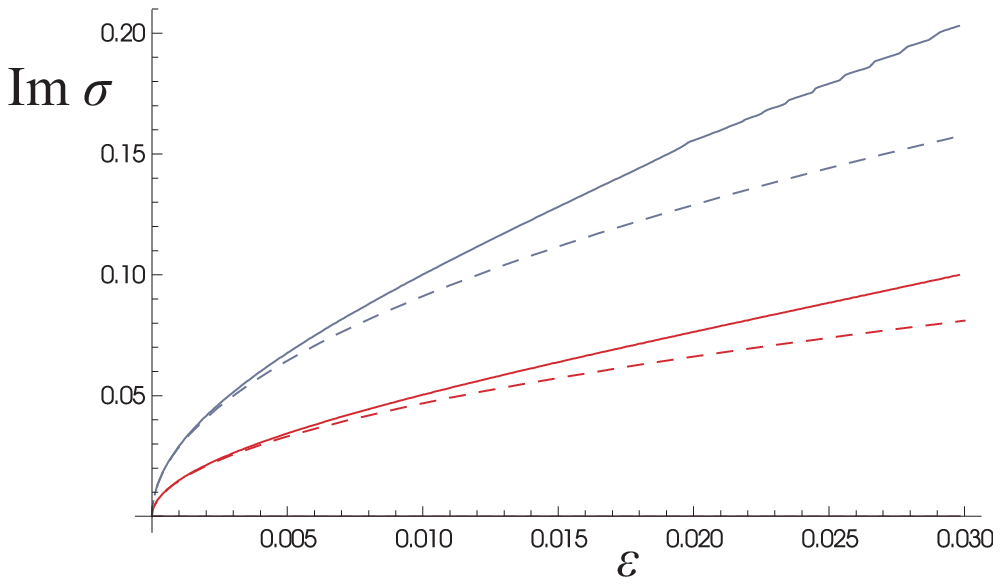}&\includegraphics[height=3.5cm]{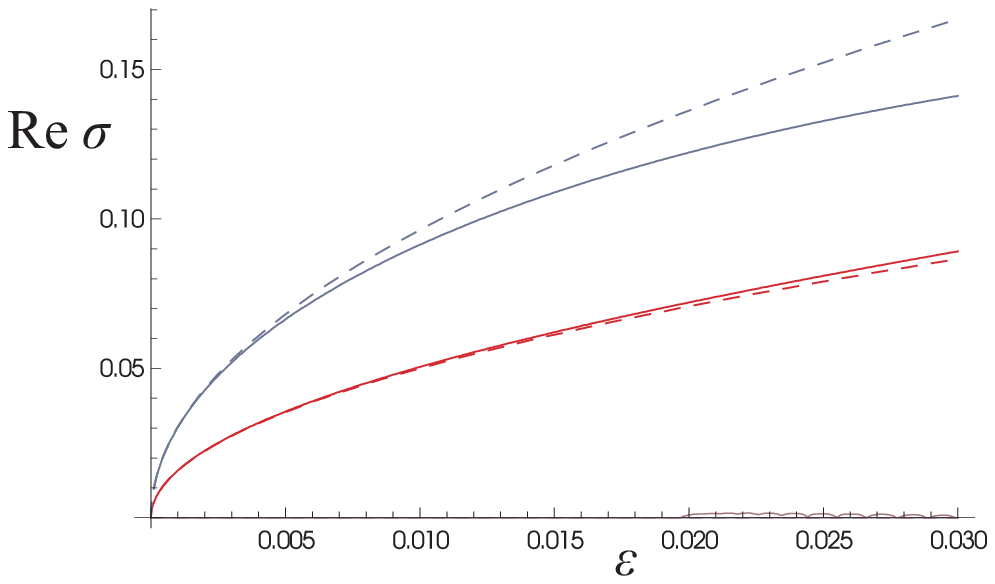}\\
\hspace{-1.5cm}(a)&(b)&(c)
		\end{tabular}
	\caption{In (a) the profile of the 5-breather corresponding to
          the $\phi_{1,2}=0, \phi_{3,4}=\pi$ configuration is
          shown  for $\ep=0.02$. In (b) and (c) the imaginary and the real parts of
          $\sigma_i$, for increasing values of $\ep$, are shown,
          respectively. The solid lines 
represent the numerically calculated values, while the dashed ones 
stem from the numerical calculation of the eigenvalues of the matrix {\bf Z}.}
	\label{fig:5site_in_in_out}
\end{figure}

Our methodology can be numerically applied (through the simple numerical
calculation of the eigenvalues of a $n \times n$ matrix), even when we
cannot analytically 
calculate the eigenvalues of $\bf E$ ($\bf Z$). In this case we can
still numerically 
calculate the eigenvalues of $\bf Z$ and get the $\cal{O}(\sqrt{\ep})$
prediction from 
(\ref{s_approx}). In order to demonstrate this, we consider five
central sites, so 
there exist four independent (relative angles) $\phi$. A
representative configuration 
is $\phi_1=\phi_2=0$ and $\phi_3=\phi_4=\pi$ which results in an
unstable multibreather 
with $\sigma_{1,2}\in \mathbb{R}$ and $\sigma_{3,4}\in\mathbb{I}$, as
expected from our main theorem above 
(see e.g. figure \ref{fig:5site_in_in_out}). 

\section{Conclusions}
In the present paper, we proved a linear stability criterion for
$n$-site multibreathers. 
This result generalizes the ones acquired in \cite{koukichtstab},
which can determine the 
stability only for configurations up to three ``central''
oscillators, while it proves a counting result concerning the number of real and imaginary characteristic exponents of the breather, which is similar to the claim stated in \cite{arcetal}. In addition, our approach provides an
$\cal{O}(\sqrt{\ep})$ estimate 
of the characteristic exponents of the multibreather
solution. Finally, the numerical 
simulations showed that our estimate is accurate for small values of
$\ep$, while it 
diverges for larger values of the coupling constant, which is
naturally expected, since for 
this range of values the higher order terms of the expansion of the
characteristic 
exponents become significant.

It would be especially interesting to extend these considerations
to higher dimensional settings, to obtain a systematic
characterization
of vortex solutions and their stability, in square, as well as
non-square
geometries. Such efforts are currently in progress \cite{kouketal} and will be
reported in future publications.

\appendix

\section{Calculation of $\avh$}\label{avh}
The average value of $H_1$ is defined as
\[\avh=\frac{1}{T}\oint H_1\ud t,\]
where the integration is performed along the unperturbed periodic orbit. Since in the anti-continuous limit $\ep=0$ the only moving oscillators are the ``central'' ones, $H_1$ becomes
\[H_1=\sum_{i=0}^n x_i^2-\sum_{i=1}^{n} x_{i} x_{i-1}.\]
Therefore, only the mixed terms of $H_1$ interest us since, as we can
easily conclude
 following the procedure below, the integration of the square terms over a period provides constant terms, i.e. terms independent of $\phi_i$. We define $I_i=\int_0^Tx_{i}x_{i-1}\ud t$, so 
\beq \avh=-\frac{1}{T}\sum_{i=1}^{n}I_i+(\mathrm{independent\ of\ \phi_i \ terms}).\label{h1}\eeq
Since at the anti-continuous limit the motion of the oscillators can be described by (\ref{fourier}) we get, by dropping the constant term of the fourier series, for $I_1$  
$$I_1=\int_0^Tx_1x_0\ud t=\int_0^T\sum_{m=1}^\infty\sum_{s=1}^\infty
A_{m}(J_1)\cos(mw_1)A_{s}(J_0)\cos(sw_0)\ud t=$$
$$\quad=\sum_{m=1}^\infty\sum_{s=1}^\infty
A_{m}(J_1)A_{s}(J_0)\int_0^T\cos[m(\w t+w_{10})]\cos[s(\w t+w_{00})]\ud
t=$$
$$\begin{array}{rl}=\ds\sum_{m=1}^\infty\sum_{s=1}^\infty
\frac{A_{m}A_{s}}{2}&\ds\left\{\int_0^T\cos[(m+s)\w
t+(mw_{10}+sw_{00})]\ud t\right.\\
&\ds\left.+\int_0^T\cos[(m-s)\w t+(mw_{10}-sw_{00})]\ud
t\right\}.\end{array}
$$
Without loss of generality, we can impose that $m, s>0$. Then, the only terms that survive are the ones with $m=s$, so we get
$$I_1=\sum_{m=1}^\infty\frac{A_{m}^2}{2}\int_0^T\cos[m(w_{10}-w_{00})]\ud t=\sum_{m=1}^\infty\frac{A_{m}^2}{2}\int_0^T\cos[m(w_{1}-w_{0})]\ud t=
\sum_{m=1}^\infty\frac{TA_m^2}{2}\cos m \phi_1.$$
So, by (\ref{h1}) we get
$$\la H_1\ra=-\frac{1}{2}\sum_{i=1}^{n}\sum_{m=1}^\infty A_m^2\cos m \phi_i.$$

\section{calculation of the $\ds\frac{\pa H_0}{\pa I_i}$ terms}\label{apb}

By taking the inverse transformation of (\ref{transformation}) we get for the $J_i$'s 

\beq\begin{array}{rcl}
J_0&=&A-I_1\\
J_1&=&I_1-I_2\\
J_2&=&I_2-I_3\\
&\vdots& \\
J_{n-1}&=&I_{n-1}-I_n\\
J_n&=&I_{n}
\end{array}\label{inversetransformation}\eeq

Since the integrable part of the Hamiltonian is written by definition 
as $H_0=H_0(J_0\ldots J_n)$ and consequently $H_0=H_0(I_1\ldots I_n)$ we get by using (\ref{inversetransformation})
$$\frac{\pa H_0}{\pa I_i}=\frac{\pa H_0}{\pa J_{i-1}}\frac{\pa J_{i-1}}{\pa I_i}+\frac{\pa H_0}{\pa J_{i}}\frac{\pa J_{i}}{\pa I_i}=-\frac{\pa H_0}{\pa J_{i-1}}+\frac{\pa H_0}{\pa J_{i}}=-\w_{i-1}(J_{i-1})+\w_{i}(J_{i}),$$
where the last equality holds because $H_0$ is separable being the sum of one degree of freedom Hamiltonians. So, every frequency depends only on the corresponding action.

So, we have 
$$\frac{\pa^2 H_0}{\pa I_i^2}=-\frac{\pa}{\pa I_i}\frac{\pa H_0}{\pa J_{i-1}}+\frac{\pa}{\pa I_i}\frac{\pa H_0}{\pa J_{i}}=-\frac{\pa^2 H_0}{\pa J_{i-1}^2}\frac{\pa J_{i-1}}{\pa I_i}+\frac{\pa^2 H_0}{\pa J_{i}^2}\frac{\pa J_{i}}{\pa I_i}=\frac{\pa^2 H_0}{\pa J_{i-1}^2}+\frac{\pa^2 H_0}{\pa J_{i}^2}$$
and for $\w_i=\w$ we get
$$\frac{\pa^2 H_0}{\pa I_i^2}=2\frac{\pa \w}{\pa J}.$$
Using the same arguments we get

$$\frac{\pa^2 H_0}{\pa I_{i+1} \pa I_i}=-\frac{\pa}{\pa I_{i+1}}\frac{\pa H_0}{\pa J_{i-1}}+\frac{\pa}{\pa I_{i+1}}\frac{\pa H_0}{\pa J_{i}}=\frac{\pa^2 H_0}{\pa J_{i}^2}\frac{\pa J_{i}}{\pa I_{i+1}}=-\frac{\pa^2 H_0}{\pa J_{i}^2}$$
which can be written as
$$\frac{\pa^2 H_0}{\pa I_{i+1} \pa I_i}=-\frac{\pa \w}{\pa J}.$$

\section{Expansion of the eigenvalues of $\bf E$}\label{oe2}
Let
\[{\bf E}=\left(
\begin{array}{c|c}
\bf O&\bf B\\
\hline
\bf C&\bf O
\end{array}
\right)\]
be the stability matrix, with eigenvalues $\sigma_i$. 
The eigenvalue problem for this matrix can be rewritten as
\[|{\bf BC}-\sigma^2\bf I|=0\]
or
\beq|{\bf BC}-\chi\bf I|=0\label{ap1_con1}.\eeq
Using the expansions 
$\chi_i=\chi_{0i}+\ep\chi_{1i}+\ep^2\chi_{2i}$, $\bf B=\ep B_1$ and $\bf
C=C_0+\ep C_1$,
 we get
$$|\ep{\bf B_1C_0}+\ep^2{\bf B_1C_1}-(\chi_0+\ep\chi_1+\ep^2\chi_2){\bf I}|=0.$$
Since this relation must hold in the limit $\ep\rightarrow 0$, we get 
$$|\chi_0{\bf I}|=0\Rightarrow \chi_{0i}=0\quad i=1\ldots n.$$
Hence, condition (\ref{ap1_con1}) becomes
$$|\ep{\bf B_1C_0}+\ep^2{\bf B_1C_1}-(\ep\chi_1+\ep^2\chi_2){\bf I}|=0$$
or
$$\left|\ep [{\bf B_1C_0}-\chi_1{\bf I}+\ep({\bf B_1C_1}-\chi_2{\bf I})]\right|=0$$
or
$$\ep^n\left|{\bf B_1C_0}-\chi_1{\bf I}+\ep({\bf B_1C_1}-\chi_2{\bf I})\right|
=0.$$
For $\ep\neq0$, this becomes
\beq\left|{\bf B_1C_0}-\chi_1{\bf I}+\ep({\bf B_1C_1}-\w\chi_2{\bf I})\right|=0.\label{divided}\eeq
However, once again, 
this condition must hold for the $\ep\rightarrow0$ limit which reads
$$|{\bf B_1C_0}-\chi_1{\bf I}|=0.$$
So, $\chi_{1i}$ depend only on $\bf B_1$ and $\bf C_0$, while $\bf
C_1$ only affects
the higher order terms.
Since
$$\sigma^2=\ep\chi_1+\ep^2\chi_2$$
$$\sigma=\pm\sqrt{\ep\chi_1}\sqrt{1+\frac{\ep\chi_2}{\chi_1}+\ldots}$$
$$\sigma=\pm\sqrt{\ep\chi_1}(1+\frac{\ep\chi_2}{2\chi_1}+\ldots)$$
hence, up to terms ${\cal O}(\ep)$, the eigenvalues of $\bf E$ are determined by $\bf B_1$ and $\bf C_0$, while the influence of $\bf C_1$ moves to terms of ${\cal O}(\ep^{3/2})$.

\section{Positive and negative eigenvalues of $\bf Z$}\label{ap_sign}
For reasons of completeness, we also present a proof of the fact that
the number of positive 
eigenvalues of $\bf Z$ ($z_i$) equals the number of positive $f_i$s
and the number of 
negative $z_i$ equals the number of negative $f_i$s. This can be done
by induction, directly
following the steps of Appendix C of \cite{sandstede}.

First we define ${\bf Z}_n$ as
\beq{\bf Z}_n=\left(\begin{array}{cccccc}
2f_1&-f_1&0 & & \\
-f_2&2f_2&-f_2&0  &  \\
 &\ddots&\ddots&\ddots& \\
  &0 &-f_{n-1}&2f_{n-1}&-f_{n-1}\\
 & &0 & -f_n&2f_n
\end{array}\right).\eeq
The determinant of ${\bf Z}_n$ is given \cite{sandstede} by
$$\mathrm{det} {\bf Z}_n=\mathrm{det}\left(\begin{array}{ccccc}
2f_1&0& & & \\
-f_2&\frac{3}{2}f_2&0& & \\
 &\ddots&\ddots&\ddots& \\
 & &-f_{n-1}&\frac{n}{n-1}f_{n-1}&0\\
 & & &-f_n&\frac{n+1}{n}f_n
\end{array}\right)=(n+1)\prod_{i=1}^nf_i.$$

The claim holds for $\mathbf{Z}_i$ with $i=1, 2, 3$. Let's assume that it holds for $\mathbf{Z}_{n-1}$. We will examine if it holds for $\mathbf{Z}_n$. Note that, we consider only the case $f_i\neq0$, since in order to have $f_i=0$, special symmetry conditions should hold.

Let's consider $f_i\neq0$ for $i=1\ldots n-1$. We define
$\tilde{f}=(f_1\ldots f_{n-1})$ and $\widehat{f}=(\tilde{f},\epsilon)$
with $\epsilon \in \mathbb{R}$, then the eigenvalues
$z_1(\epsilon)\ldots z_n(\epsilon)$ of $\mathbf{Z}_n$ are $C^1$ in
$\epsilon$; see e.g. \cite{yac}.

Consider $\epsilon=0$ first. Since $f_i\neq0$ for $i=1,\ldots, n-1$ we have that $z_1(0),\ldots, z_{n-1}(0)\neq0$. In addition

$$\prod_{i=1}^{n-1}z_i(0)=n\prod_{i=1}^{n-1}f_i\neq0\Rightarrow\mathrm{sign}\left(\prod_{i=1}^{n-1}z_i(0)\right)=\mathrm{sign}\left(\prod_{i=1}^{n-1}f_i\right)\neq0$$
and $z_n(0)=0$ since the last row of ${\bf Z}_n$ vanishes. For $\epsilon\neq0$ we have
$$\prod_{i=1}^{n} z_i(\epsilon)=(n+1)\epsilon\prod_{i=1}^{n-1}f_i$$
so, for small $\epsilon$ it is
$$\mathrm{sign}(z_n(\epsilon))=\mathrm{sign}(\epsilon).$$
But, 
$$\det{\bf Z}_n=(n+1)\epsilon\prod_{i=1}^{n-1}f_i\neq0$$
is valid for every $\epsilon\neq0$, so no eigenvalue can change sign as long as $\epsilon$ is nonzero. Consequently the claim holds for $\mathbf{Z}_n$ which coincides with $\mathbf{Z}$, so the lemma is proven.\\[7pt]
\ack The authors would like to thank Dmitry Pelinovsky for a
  critical read of the manuscript and numerous helpful remarks.
PGK also gratefully acknowledges support from
NSF-DMS-0349023, NSF-DMS-0806762 and the
Alexander von Humboldt Foundation.

\bibliographystyle{unsrt}
\bibliography{mybib}

\end{document}